# Higgs Boson Production at the LHC

M. Y. Hussein*

*Department of Physics, College of Science, University of Bahrain
P.O. Box 32038, Kingdom of Bahrain

One of the major goals of the Large Hadron Collider (LHC) is to probe the electroweak symmetry breaking mechanism and the generation of the masses of the elementary particles. We review the physics of the Higgs sector in the Standard Model. The main production channels of the Higgs production at the LHC are reviewed. The prospects for the discovering the Higgs particles at the LHC and the study of their fundamental properties are summarized.

## 1. Introduction

The Higgs mechanism is the cornerstone in the electroweak sector of the Standard Model [SM]. The fundamental particles, leptons, quarks and gauge particles, acquire the masses through the interaction with a scalar field [1,2]. To accommodate the well–established electromagnetic and weak phenomena, this mechanism requires the existence of at least one scalar field. After absorbing three Goldstone bosons, one degree of freedom is left over which corresponds to a scalar particle "Higgs boson". The properties of the Higgs boson, decay widths and the production mechanisms, can be predicted if the mass of the particle is fixed.

Although the Higgs boson is an essential ingredient in the Standard Model, but it has not yet been observed. After the end of LEP program, the Higgs search will be carried out at hadron colliders. If nature had chosen the most elegant and symmetric mathematical construct for the SM, all fundamental particles would have zero mass. Experiments show that this is not the case: all particles have a small mass.

The Standard Model has no mechanism that would account for any of these masses, unless we supplement it by adding additional fields, of a type known as scalar fields. One of the greatest achievements of twentieth century theoretical physics was the discovery of how to incorporate particle masses into the Standard Model theory without spoiling the symmetry and mathematical consistency. The most straightforward way to do that is through what is known as "Higgs mechanism" [1, 2]

The discovery of the Higgs boson would solve the mystery of how particles get their masses. The spontaneous breakdown of the weak-electromagnetic gauge group can be accomplished in only one known way "Higgs fields".

The Higgs is elusive and plays a negligible role in low energy phenomenology because it couples with particles according to their masses, and so couples very weakly to the lepton and quark constituents of ordinary matter.

The Higgs bosons are produced through a variety of mechanism at the LHC. Each of them makes use of the preference of the SM Higgs which couples to the heavy particles, either massive vector boson or massive quark. The LHC is expected to find the Higgs boson for all masses less than around 800 GeV. The production cross sections are large and the theoretical predictions are well understood, with all important channels known to at least to next-to-leading order accuracy.

Figure 1 shows the production cross section of the Higgs boson as a function of Higgs mass at the LHC.

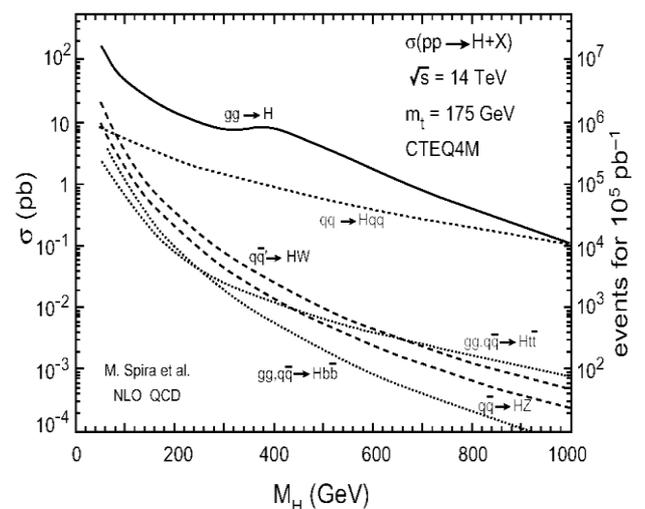

Figure 1: The production cross section and number of events for $10^5$ pb$^{-1}$ of the Higgs boson as a function of the Higgs mass at the LHC.

The gluon-gluon fusion through a heavy quark loop is the dominant production mechanism and the largest mechanism at hadron colliders, but again the largest decay for light Higgs boson into bottom quarks, has an overwhelming QCD background. At the Tevatron Run II, it leads to about 65% of the total cross section in the range 100–200 GeV. At the LHC gg fusion dominates over the other production channels for a light Higgs [3].

The vector boson fusion process is the second most dominant production mode at the LHC. It typically takes to 10% of the total Higgs cross section for low masses and up to 50% for a very Higgs boson. The vector boson fusion channel qq $q\bar{q} \rightarrow qqVV \rightarrow qqH$, which is not important at the Tevatron, is useful for the Higgs discovery over a large range of Higgs mass at the LHC [4]. By tagging the forward jets associated with the Higgs production, the background can be significantly reduced. The QCD corrections have been computed within the structure function approach and they increase the cross section of about 10% both at the Tevatron and at the LHC.

The most promising channel for the Higgs discovery is the associate production of Higgs with vector boson $q\bar{q} \rightarrow V^* \rightarrow VH$ for mass of Higgs < 135 GeV, where the bb decay is dominant [5]. This is due to the possibility to trigger on the leptonic decay of the vector boson. The QCD corrections are the same as for Drell–Yan and increase the cross section of 30% at the Tevatron Run II and of 25–40% at the LHC.

The associated production of a Higgs with a pair of t quarks at the LHC, $q\bar{q}$, $gg \rightarrow Ht\bar{t}$ will play a very important role in the low Higgs boson mass range, both for the discovery and for precision measurements of the Higgs boson couplings. This process will provide a direct measurement of the top quark Yukawa coupling [6]. This production with $H \rightarrow b\bar{b}$, could be observed at the LHC with sufficient luminosity.

The associated production of a Higgs with a pair of b quarks at the LHC, has small cross section which is due to small size of Yukawa coupling. In some extensions of the SM, such as the MSSM, the Yukawa coupling of the b-quark can become strongly enhanced, the associate production of a Higgs with a pair of b quarks can dominate over the other production channels and this production can be a significant source of Higgs boson.

The final states most suitable for discovery at the LHC vary depending on the branching ratios, shown in figure 2, which are a function of the Higgs mass, and the relevant backgrounds [7]. For mass of Higgs less than $2M_W$ the dominant decay mode is through bb. However, due to enormous QCD background, this channel is only considered in the ttH final state where handles exist for the rejection of this background. If the Higgs mass is large enough to make the $WW$ and $ZZ$ modes kinematically accessible, the H→WW final states over a very large mass range, as is the $H \rightarrow ZZ \rightarrow l\bar{l}l\bar{l}$ four leptons, the later is commonly to us the 'golden mode' as with four leptons in the final state is easy to trigger and allows for full reconstruction of the Higgs mass.

Both CMS and ATLAS at the LHC [8] conducted extensive fully simulated Monte Carlo studies to determine the experimental viability of all these channels.

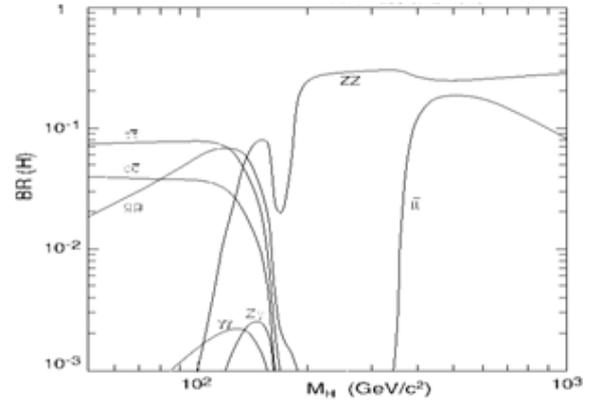

Figure 2. Branching ratios for standard model Higgs decays

A more comprehensive and complete account can be found elsewhere [8].

The purpose of this work is to calculate the expected cross section for Higgs boson at LHC energy, for both single and double parton scattering mechanism.

This paper is organized as follows. In the following section we discuss the double parton scattering mechanism for Higgs boson production. In sction 3 we calculate all possible production for Higgs boson at LHC energy. Our summery and discussions on the Higgs boson production are presented in Section 4.

## 2. Double Parton Scattering Mechanism

Many features of high energy inelastic hadron collisions depend directly on the parton structure of hadrons. The inelastic scattering of nucleons need not to occur only through a single parton-parton interaction and the contribution from double parton collision (DP) collisions can be significant. A schematic view of a double parton scattering events in a $p\bar{p}$ interaction is shown in Fig. 3

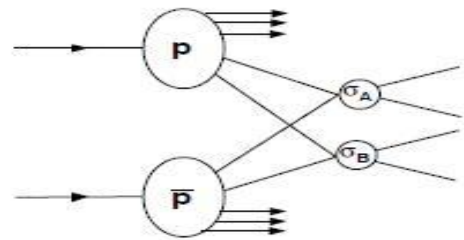

Figure 3. Diagram of a double parton scattering event.

The rate of events with multiple parton scattering depends on how the partons are special distributed within the nucleon. So at high energies and due to large flux in particular at the LHC, another type of scattering phenomenon is expected to contribute the cross section besides to single scattering: proton-proton collisions with multiple parton interactions [9]. Thus for Higgs boson production there would be two computing mechanisms: single parton scattering and double (multiple) parton scattering featuring two Drell-Yan processes happening simultaneously [10,11, 12].

The purpose of the present work is to calculate the expected cross sections for Higgs boson production at the LHC, from both single and double parton scattering mechanisms.

Multiple parton interaction processes, where different pairs of partons have hard scattering in the same hadronic at high energies because of the growing flux of partons.

The possibility of hadronic interactions with double parton scattering collisions was foreseen on rather general ground long ago [5]. The AFS experiment at the CERN ISR has claimed evidence for the double scattering in pp collisions [6]. The collider detector at FERMILAB CDF has observed the process by looking at final state with three mini-jets and one photon [7]. The multiple parton scattering occurs when two or more different pairs of parton scatter independently in the same hadronic collision. In principle double scattering probes correlation between partons in the hardon in the transverse plane, which provides extra additional information on hardon structure. The many parton distributions considered in the multiple parton scattering are correlated. These parton distribution depend on the fractional momenta of all the interacting partons $x$ and on their distance in transverse space $\hat{b}$. The inclusive cross-section of a double parton scattering can be expressed, as:

$$\sigma_{DS} = \Sigma \int dx_1 \, dx_2 \, dx_3 dx_4 d \, \Gamma_A(x_1, x_2; \hat{b})$$
$$\Gamma_B(x_3, x_4; \hat{b}) \, \hat{\sigma}^a(x_1, x_3) \hat{\sigma}^b(x_2, x_4) d^2 b \quad (1)$$

Where $\hat{\sigma}^a(x_1, x_3), \hat{\sigma}^b(x_2, x_4)$ are two partonic cross sections and $\Gamma_A(x_1, x_2; \hat{b})$ represents the two body parton distribution function with frictional momenta $x_1, x_2$ and transverse relative distance with in the hadron $\hat{b}$ [7]. These correlations can be assumed to be negligible if a scattering event is characterized by high centre-of-mass energy. Then the two-body parton distribution functions factorize.

$\Gamma(x_1, x_2; \hat{b}) = f(x_1) f(x_2) F(\hat{b})$, $f(x)$ is the usual one body parton distribution and $F(\hat{b})$ discribes the distribution of parton in the transverse plane. With these assumptions the cross-section for a double collision leads in the case of two distinguishable parton interactions to the simplest factorized expression:

$$\sigma_{DS} = \frac{\sigma^a \sigma^b}{\sigma_{eff}} \quad (2)$$

Here $\sigma^a$ represents the single scattering cross section. If the two interactions are indistinguishable, double counting is avoided by replacing

$$\sigma_{DS} = \frac{\sigma 2_{SS}}{\sigma_{eff}} \quad (3)$$

Here $\sigma 2_{SS}$ represents the single scattering cross section

$$\sigma_{SS} = \Sigma \int dx_a \, dx_b f_i(x_a) f_i(x_b) \hat{\sigma} ij \to a \quad (4)$$

With $f_i(x_a)$ the standard parton distribution and $\hat{\sigma} ij \to a$ represents the sub-process cross section. The parameter $\sigma_{eff} = 1/\int d^2 \hat{b} F(\hat{b})$, is the effective cross section and it enters as a simple proportionality factor in the integrated inclusive cross section for a double parton scattering $\sigma_{DS}$. The value of $\sigma_{eff}$ represents therefore the whole output of the measure of the double parton scattering process, which on the other hand has shown to be in agreement with the available experimental evidence. The experimental value measured by CDF yields $\sigma_{eff} = 1.45 \pm 1.7^{+1.7}_{-2.3}$ mb [11].

## 3. Cross-Section Results

For the production of Higgs boson at the LHC collider, we evaluate the fully cross section for the simplest spontaneous broken gauge boson involving exactly one physical SM Higgs boson via several mechanisms.

There are essentially four mechanisms for the single production of the SM Higgs boson at th LHC; Some Feynman diagrams are shown in Fig. 4.

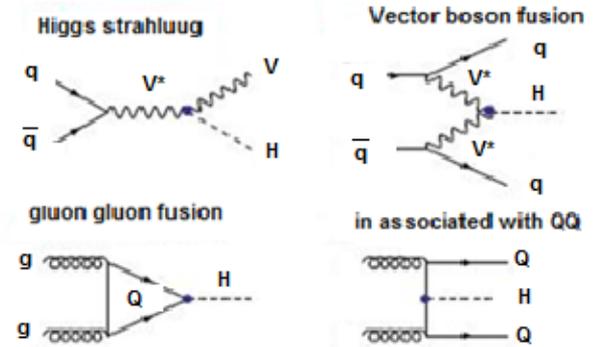

Figure 4. The production mechanisms for SM Higgs bosons at hadron colliders.

The total cross sections for leading order sub-presses Higgs boson production for various channels, are evaluated using GS09 parton distributions [13], the packages Mad-graph and HALAS, and the integration was performed by VEGAS for the LHC with energy $\sqrt{s}$ = 14 TeV and $m_H$ = 120 GeV are displayed in Table 1.

A comparison of the cross section for single and double parton scattering is shown in Table 1.

| Higgs Production | $\sigma_{SS}$ (pb) | $\sigma_{DS}$ (pb) |
|---|---|---|
| $q\bar{q} \to H$ | 50 | ---- |
| $q\bar{q} \to Hqq$ | 6.0 | 2.50 |
| $q\bar{q} \to WH$ | 2.7 | 0.19 |
| $q\bar{q} \to ZH$ | 1.6 | 0.17 |
| $q\bar{q} \to t\bar{t}H$ | 0.67 | 0.20 |
| $q\bar{q} \to b\bar{b}H$ | 1.5 | 0.67 |

Table 1. The total cross-section for Higgs boson production at $\sqrt{s}$ = 14 TeV.

The Higgs boson will be predominantly produced via gluon-gluon fusion. For the Higgs masses, such that $m_H$ > 100 GeV/c$^2$, the second dominant process is the vector boson fusion. Associated production mode, where the Higgs is produced via $q\bar{q} \to HW, HZ, t\bar{t}H, b\bar{b}H$, has smaller cross-sections. The presence of a W, Z, top or bottom quarks alongside the Higgs, or high-p$_T$ high-$\eta$ jets from VBF, allow for triggering on events with Higgs in invisible final states.

## 4. Summary

With the turn on of the LHC, particle physics will enter a new era of electroweak physics. There are three possibilities for the Higgs sector. First, the Higgs could discover with the SM-like with mass consistent with the electroweak precision, the second is that the Higgs boson is discovered with SM-like but with a mass inconsistent with the electroweak observables. Finally, it is possible that no Higgs boson will be discovered. In this case, it is possible that new particles outside the Higgs sector will be discovered.

The main production channels of the Higgs boson at LHC are briefly reviewed and recent developments in the calculation are discussed.

In this work we have computed the most promising channel to detect the production of the SM Higgs particle at the LHC with a mass below about 135 GeV, where the Higgs decays into $b\bar{b}$ final states is dominant.

The production of the Higgs boson, in the intermediate mass range, namely the final state with a $b\bar{b}$ pair and with isolated lepton, is affected by a sizable background due to double parton collision process.

If the Higgs boson decay into two W bosons and subsequently into two leptons and two neutrinos ($H \to WW \to l\nu l\nu$) is expected to be the main discovery channel for the intermediate Higgs boson mass range, between $2m_W$ and $2m_Z$. The signature of this decay is characterized by two leptons and high missing energy.

Motivations for studying the $WH$ associated production, with a subsequent decay of the Higgs boson into a $W$ pair are twofold. First, as already mentioned, this channel is one of the few possibilities close to the $WW$ resonance. Second, the corresponding Feynman diagrams contain the $g_{HWW}$ coupling constant twice, which could be precisely measured. Since three $W$ bosons are produced in this process, the final state is characterized by six fermions in addition to soft remnants from the protons. The three lepton channel provides a clean signature and has an interesting signal over background ratio.

Although the double parton scattering background cross section is a decreasing function of the invariant masses of the $b\bar{b}$ pair, the relatively large value of the invariant mass required to the $b\bar{b}$ pair to be assigned to the Higgs decay is not large enough, at LHC energies, to allow one to neglect the double parton scattering background.

However, in the hardon collider environment, the large QCD backgrounds may cause the observation impossible for those events if one or more of the gauge bosons decay hadronically. Individual channels with hadronic decays should be studied on case by case.

It is useful to point some of the inherent uncertainties which affect the final results. The most significant are: (i) the lack of precise knowledge of the parton distribution at small $x$, which is important for the intermediate mass Higgs and (ii) the effect of unknown higher order perturbation QCD corrections.

## Acknowledgements

I would like to thank W. J. Stirling and D. Treleani for very useful discussions. I would also thank the staffs at the Theory Group at CERN, for scientific collaboration and hospitality.

## References

[1] F. Englert and R. Brout, Phys. Rev. Lett. 13, (1964) 321; P. W. Higgs, Phys. Lett. B12, (1964) 132; P. W. Hi\ggs, Phys. Rev. Lett. 13, (1964) 508.
[2] A. Djouadi, hep-ph/0503172.
[3] S. Catani, D. de Florian and M. Grazzini, Nucl. Phys. B596, (2001) 299.
[4] M. Spira, Fortsch. Phys. 46, (1998) 203.
[5] M. Y. Hussein, Nucl. Phys. Proc. Suppl. 174, (2007) 55; arXiv: 0710.0203[hep-ph].
[6] W. Beenakker, S. Dittmaier, M. Kramer, B. Plumper, M. spira, P. Zerwas, Phys. Rev. Lett. 86, 201805 (2001).
[7] M. Spira, http://people.web.psi.ch/spira/.
[8] CMS Physics TDR, CERN-LHCC 2006-021;
ATLAS Physics TDR, CERN-LHCC 99-14/15.
[9] A. del Fabro and D. Treleani, Phys. Rev. D61, (2000) 077502.
[10] M. Drees and T. Han, Phys. Rev. Lett. 77, (1996) 4142.
[11] F. Abe *et. al.* [CDF Collaboration], Phys. Rev. D 58, (1997) 3811; V. M. Abazov *et al* [CDF Collaboration], Phys. Rev. D 81, (2010) 052012.
[12] Abazov *et. al.* [FERMILAB-PUB-09-644-E], arXiv:0912.5104v2[hep-ex]
[13] J. R. Gaunt and W. J. Stirling, JHEP 1003, (2010) 005 [arXiv:0910.4347 [hep-ph]].